\documentclass[prl,twocolumn,aps,amssymb]{revtex4}
\usepackage{epsfig}
\usepackage{subfigure}
\usepackage{bbm}
\begin{document}

\title{Enhancement of Noise-induced Escape through the Existence of a Chaotic Saddle}
\author{Suso Kraut$^{(1,2)}$ and Ulrike Feudel$^{(2)}$}
\affiliation{$^{(1)}$ Institut f\"ur Physik, Universit\"at Potsdam,
Postfach 601553,D-14415 Potsdam, Germany \\
$^{(2)}$ ICBM,
Carl von Ossietzky Universit\"at, PF 2503, 26111 Oldenburg, Germany}
\date{\today}

\begin{abstract}

We study the noise-induced escape process in a prototype dissipative
nonequilibrium system, the Ikeda map. In the presence of a chaotic saddle
embedded in the basin of attraction of the metastable state, we find the novel
phenomenon of a strong enhancement of noise-induced escape. This result
is established by employing the theory of quasipotentials. Our finding is
of general validity and should be experimentally observable.

PACS number 05.45+b\\
\end{abstract}

\maketitle
\begin{figure}[htb]
\begin{center}
\epsfig{file=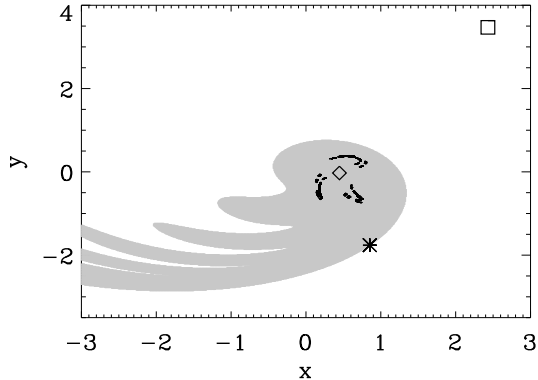,width=7cm,height=5cm}
\caption{Grey dots represent the basin of attraction for the fixed point
marked with a $\Diamond$ for $\eta = 4.1$. The other fixed point is also
depicted ($\Box$). The chaotic saddle is shown with black dots and the
saddle point on the basin boundary is marked by {\Large $\ast$}.}
\label{basin_ikeda}
\end{center}
\end{figure}
\begin{figure}[htb]
\begin{center}
\epsfig{file=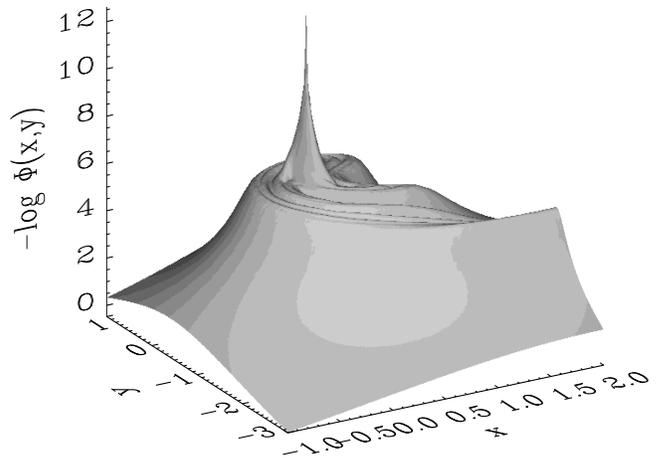,width=8.5cm,height=6cm}
\caption{Quasipotential $\Phi(x,y)$ for the Ikeda map with $\eta = 4.1$ on a
$300 \times 300$ grid. The single peak corresponds to the fixed point.
Also, an extended plateau at $-\log \Phi(x,y) \approx 5.0$ is visible,
caused by the chaotic saddle. }
\label{QP_ikeda_4.1}
\end{center}
\end{figure}
\begin{figure}[!htb]
\begin{center}
\epsfig{file=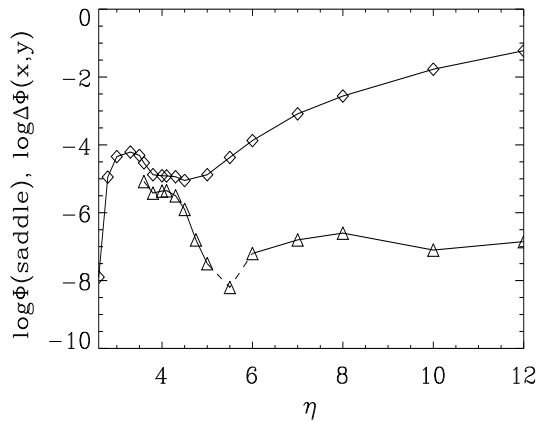,width=7cm,height=5.5cm}
\caption{Minimal escape energy ($\Diamond$) and height of the saddle
plateau  ($\triangle$) in logarithmic scale versus $\eta$.}
\label{QP_escape_saddle}
\end{center}
\end{figure}
\begin{figure}[!htb] 
\begin{center}
\epsfig{file=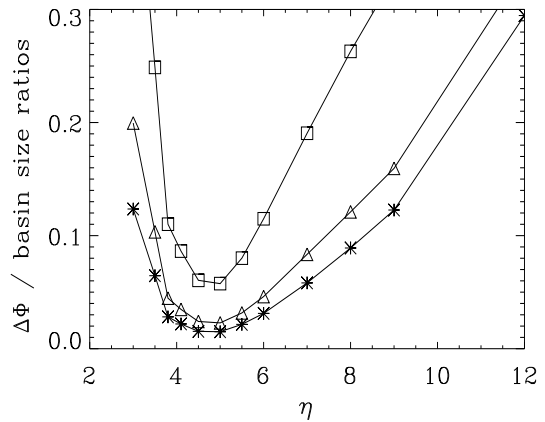,width=7cm,height=5.5cm}
\caption{Average quasipotential height (= escape energy) divided by the ratios
of the sizes of the basins of attraction (\# initial conditions A / (A + B)).
The curves correspond to a smaller frame of reference from top to bottom,
with the values $x \in [-10.0,10.0], y \in [-10.0,10.0]$ (marked with $\Box$),
$x \in [-5.0,5.0], y \in [-7.0,7.0]$ (marked with $\triangle$),
$x \in [-3.0,2.0], y \in [-3.5,4.0]$
(marked with {\LARGE $\ast$}), respectively.}
\label{ikeda_escape_vs_basin}
\end{center}
\end{figure}
\begin{figure}[!htb]
\begin{center}
\epsfig{file=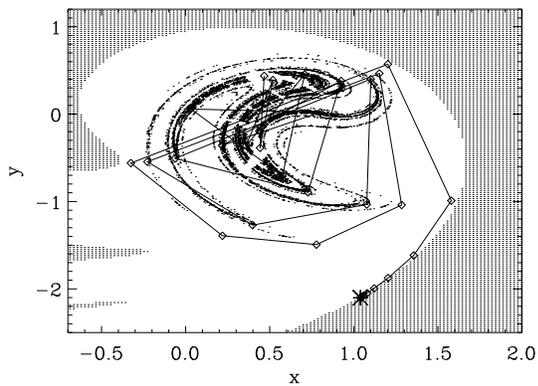,width=7cm,height=5cm}
\caption{Escape path and chaotic saddle for $\eta = 5.0$.
The saddle point on the basin boundary is shown as {\Large $\ast$}.}
\label{escape_path}
\end{center}
\end{figure}

Since the seminal treatment of the noise-induced escape problem by Kramers
\cite{Kramers:1940}, major progress has been made by Onsager and
Machlup. They realized that the escape process consists of large
fluctuations, which are very rare, and that the trajectory peaks sharply
around some optimal (most probable) escape path \cite{Onsager:1953}.
Thus, despite the stochastic nature of the escape process, the escape path
is of almost deterministic nature, as other paths than the most
probable one have an exponentially smaller probability. That theory
was derived for a small noise level  $\delta \rightarrow 0$. A review
on noise-induced escape in equilibrium systems and the most important
recent advancements is given in \cite{Hanggi:1990}.
\\
In the last decade it has been realized, that systems that are not in thermal
equilibrium or are lacking the property of detailed balance, can give rise
to a large variety of interesting phenomena in the noise-induced escape problem.
Only recently experiments on this problem have been conducted, using
Josephson junctions, electronic circuits, lasers,
and an electron in a Penning trap \cite{Vion:1996}.
Some of the most interesting novel theoretical findings include a pre-exponential
factor of the Kramers rate \cite{Maier:1992}, a symmetry breaking bifurcation
of the optimal escape path \cite{Maier:1993} and a distribution of the escape paths
originating from a cusp point singularity \cite{Dykman:1996}. Furthermore, the very
intriguing phenomenon of saddle-point avoidance has been discovered
\cite{Luchinsky:1999a}. For a fluctuating barrier the effect of resonant activation
has been theoretically predicted  \cite{Doering:1992} and experimentally confirmed 
\cite{Mantegna:1996}. Also a stepwise growth of the escape rate for short time scales
has been found \cite{Soskin:2001a}. Recently, an oscillation of the escape rate in
dependence on the friction for a multiwell potential was demonstrated \cite{Soskin:1999}.
For periodically driven systems  a number of interesting results has been obtained
as well, like a resonantly decrease in the activation energy \cite{Dykman:1997},
a logarithmic susceptibility of the fluctuation probability \cite{Smelyanskiy:1997},
time oscillations of escape rates \cite{Smelyanskiy:1999} and enhancement of escape
due to transient chaos \cite{Soskin:2001b}. 
\\
Here we report on a new mechanism of lowering the required energy for
noise-induced escape (enhancement of escape), thus a reduction of the
mean first passage time. This happens, if a chaotic saddle is embedded
in the open neighborhood of the basin of a metastable state. Then the
escape trajectory does not only pass through a single unstable periodic
orbit \cite{Khovanov:2000}. By contrast, it can pass through 
the chaotic saddle, i. e.  a geometrically strange, invariant,
non-attracting set (which is made up of an infinite number of unstable
periodic orbits). The trajectory can jump between points of the chaotic
saddle  with no additional activation energy required.
The overall lowering of the activation threshold is due to the fact,
that the escape process consists now of three subsequent steps: Firstly,
the trajectory jumps on one orbit on the chaotic saddle. Secondly, it
switches on the chaotic saddle, without need of input energy, to select the
orbit which allows the easiest escape, and thirdly it fluctuates from
that orbit to the saddle point on the basin boundary. By this mechanism,
the chaotic saddle is transformed into a dynamically relevant quantity,
whereas in noisefree systems it is only important for transient behavior.
In this way the chaotic saddle acts as a `shortcut'.
\\
Since noise-induced escape has previously been studied using
dissipative maps \cite{Beale:1989}, which allow analysis in a
straitforward way, we demonstrate our findings for the Ikeda map
\cite{Ikeda:1979}. This is an idealized model of a laser pulse i
n an optical cavity. With complex variables it has the form
\begin{equation}
z_{n+1} = a + b z_n \exp \left[ i \kappa - \frac{i \eta}{1 + |z_n|^2} \right],
\end{equation}
where $z_n = x_n + i y_n$ is related to the
amplitude and phase of the $nth$ laser pulse exiting the cavity. The
parameter $a$ is the laser input amplitude and corresponds to the
forcing of the system. The damping ($1-b$) accounts for the reflection
properties of mirrors in the cavity and measures the dissipation.
The empty cavity detuning is given
by $\kappa$ and the detuning due to a nonlinear dielectric medium by
$\eta$. The Ikeda map gives rise to rich dynamical behavior, exhibiting
for some parameters even highly multistable behavior
\cite{Feudel:1997}.
\\
We fix the parameters at $a = 0.85, b = 0.9$ and $\kappa = 0.4$ and
vary only $\eta$ in the range $2.6 < \eta < 12$. For the noiseless system
two stable states are present. One fixed point (state A, $\Diamond$ in
Fig. \ref{basin_ikeda}) undergoes a period doubling scenario and becomes
a chaotic attractor at $\eta \approx 5.5$ Another fixed point (state B,
$\Box$ in Fig. \ref{basin_ikeda}) remains a fixed point over the whole
parameter range considered. The noise-induced escape from state A is
investigated. The basin boundary separating these two stable states is
a smooth curve which is build by the stable manifold of the saddle
point C ({\Large $\ast$} in Fig. \ref{basin_ikeda}), separating the two stable states. 
With the above mentioned features of the system, there are no unusual effects 
expected in the noise-induced escape problem. However,
for a critical value of $\eta_c = 3.5686$ an additional period $3$ solution close
to the fixed point (state A) emerges, with a fractal basin boundary
between these two solutions and a chaotic saddle embedded in this fractal basin
boundary.
It is important to note, that the basin boundary between the two fixed points
remains smooth over the whole parameter range considered here.
Increasing $\eta$ further, the stable period 3 solution dissapears in a
boundary crisis, yet a chaotic saddle is {\it still} present beyond the
boundary crisis, completely embedded in the open neighborhood of the basin
of the stable solution, as can be seen in Fig. \ref{basin_ikeda} for $\eta = 4.1$.
A chaotic saddle is a geometrically strange, invariant, non-attracting set.
It is computed using the PIM-triple algorithm \cite{nusse:1989}.
We stress that it is the chaotic saddle, that has a remarkable effect on the
average escape time from the stable fixed point.
\\
To treat the problem of noise-induced escape we now employ the theory
of quasipotentials, which gives rigorous results on the influence of
noise on the invariant density and the mean first passage time.
Quasipotentials have been introduced in the mathematical literature
for time-continuous systems in \cite{Freidlin:1984} and for discrete
time ones in \cite{Kifer:1988}. For systems of physical interest, they were
first proposed in \cite{Graham:1984} and extended to systems with coexisting
attractors in \cite{Graham:1986}. Discrete systems with strange invariant
sets were for the first time treated in \cite{Graham:1991}.
Quasipotentials can be derived through a minimization procedure of
the action of escape trajectories from a Hamilton-Jacobi equation
\cite{Talkner:1987}. The action to be minimized has the form
\begin{equation}
\label{action_QP}
S_N[(z_i)_{0 \le i < N}] = \frac{1}{2} \sum_{i=0}^{N-1}[z_{i+1} - f(z_i)]^2,
\end{equation}
for the map $z_{n+1} = f(z_n) + \sigma \,\xi_n$, where $\sigma$ is the
standard deviation of the additive, Gaussian, white noise term $\xi_n$.
With appropriate boundary conditions the infimum of this action with
respect to $N$ and $i$ along a path is the quasipotential $\Phi$.
The mean first exit time is then given in analogy to Kramer's law:
\begin{equation}
\label{MFPT_QP}
\langle\tau \rangle \sim \exp \left[ \frac{\Delta \Phi}{\sigma^2}\right], 
\end{equation}
whith $\Delta \Phi$ defined as the minimal quasipotential difference
\begin{equation}
\Delta \Phi:= \inf \{ \Phi(y) - \Phi(a):a \in A, y \in \partial G \},
\end{equation}
where $A$ is the attractor and $ \partial G$ is the basin boundary.
In Fig. \ref{QP_ikeda_4.1} the quasipotential is shown for  $\eta = 4.1$.
There is a single peak corresponding to the fixed point A and a plateau region
of a practically constant quasipotential, which reflects the chaotic saddle.
\vspace{0.5cm}
Employing the quasipotential for the noise-induced escape problem, the
minimum value of $\Phi(x,y)$ on the basin boundary has to be determined.
This is exactly the minimum escape energy $\Delta \Phi(x,y)$, since the
quasipotential at the stable solution (fixed point, periodic orbit or
chaotic attractor) is zero. The point on the basin boundary, where this
happens, is generally a saddle point of the system.
\\
To quantify the escape process with the quasipotential, we plot for
various values of $\eta$ the corresponding minimal escape energy $\Delta
\Phi(x,y)$ in Fig. \ref{QP_escape_saddle}. To elucidate the role of
the chaotic saddle as the origin of an enhancement of noise-induced escape,
we also include in the plot the value of the height of the plateau in the
quasipotential. 
\\
In the framework of quasipotentials, the difference in height of the
escape energy and the saddle plateau corresponds to the distance between
the basin boundary and the chaotic saddle, whereas the height of the saddle
is related to the distance between the attractor and the saddle.
\\
The mechanism of the escape process is closely connected to the
existence of an embedded chaotic saddle. It consists of two steps, namely, a
noise-induced fluctuation from the attractor (state A) to the chaotic saddle,
and then from the chaotic saddle to the fixed point (state B) via the saddle
point on the boundary. The escape can also be incomplete, as the trajectory
may fall back from the chaotic saddle to the attractor.
In a successful escape, the chaotic saddle acts as a `shortcut', as
its presence lowers the overall escape energy. This behavior seems to
be especially pronounced if the chaotic saddle is closer to the basin
boundary than to the attractor (compare the region $3.6 \le \eta \le 4.5$
of  Fig. \ref{QP_escape_saddle}). Let us note that for $\eta = 5.5$ it is not
clear, if there exists a chaotic saddle, which is the case for all
other values of $\eta \ge 3.5686$ we have tested. The PIM-triple
method, as well as the quasipotential, yield for $\eta = 5.5$ no
conclusive result, as a chaotic saddle may exist {\it very close} to
the chaotic attractor and numerically it is very difficult to
distinguish between the two.
\newline
It is important to quantify the influence of the relative sizes of the
basins of attraction, since it is increasing with increasing $\eta$ and
we are here only interested in the change of activation energy caused by
the chaotic saddle.
The distance between the attractor and the saddle point on the boundary is
usually proportional to the relative size of the basin. Both quantities are
expected to play a role in the stability of the metastable state located in
the basin, although we are not aware of any theoretical work dealing with this
relation directly.
To compensate for the  change of escape energy caused by the increase in size
of the basin of attraction, in Fig.  \ref{ikeda_escape_vs_basin} the escape energy
is divided by the ratio of the size of the basin of attraction of state A
to the overall area (A + B) for $3$ different sections of the phase space. The
sections of the phase space decrease from top to bottom. 
The combination of the two quantities, potential height and basin size,
yields a pronounced minimum at $\eta \approx 5.0$ for all $3$ curves, thus
confirming the essential role in lowering the escape energy played by the
chaotic saddle.
The most probable escape path \cite{Onsager:1953,Freidlin:1984} for
$\eta = 5.0$ is shown, together with the chaotic saddle, in Fig.
\ref{escape_path}. For this parameter value, there is a stable period 4
solution. As can be seen, the trajectory jumps at first directly on points of
the chaotic saddle, moves secondly along points of the chaotic saddle
for some iterations, until it is thirdly transported close to the basin
boundary to the saddle point. Since the first step (from fixed point to saddle)
and the last step (from saddle to the basin boundary) are minimal in
this case, the enhancement is maximal.
Other values of the Ikeda map, where no
chaotic saddle is present, have been investigated as well. For these parameter
values the effect could not be found, and the graphs corresponding to
Fig. \ref{ikeda_escape_vs_basin} have a strictly monotonic shape.
This demonstrates that the existence of the chaotic saddle is of
crucial importance for the occurrence of the enhancement of noise-induced
escape.
\\
Moreover, the stability of a fixed point is determined by its
eigenvalues. The eigenvalues are found to be $\lambda = 0.9$ for the whole
range, where it exists. Consequently, the linear approximation
is of no relevance to the noise-induced escape problem, as its range
of validity is much smaller than the region for the escape, which is the whole
open set of the basin of attraction shown in Fig. \ref{basin_ikeda}. 
\\
To conclude, we have demonstrated the effect of enhancement of noise-induced
escape through the existence of a chaotic saddle in the open neighborhood of
the metastable state for the Ikeda map as a parameter is varied. Employing the theory
of quasipotentials, it was possible to understand this lowering of the
escape threshold. We stress that the reported mechanism of the lowering of the
escape energy is of qualitatively different nature from a recently found
effect, where also an enhancement of noise-induced escape through transient motion
(typical for chaotic saddles) has been found \cite{Soskin:2001b}. In this scenario, a
nonadiabatically, periodically driven system exhibits a facilitation
of noise-induced interwell transitions. This occurs, because the
basin boundary becomes  fractal and the distance between the two states
is effectively reduced. In the mechanism reported here we always have a
smooth basin boundary between the two states and the chaotic saddle is
embedded in the basin of one state, not in the basin boundary between
the states. The analysis of the exact escape path on the chaotic saddle,
in contrast to the case where the trajectory leaves via a single periodic
orbit \cite{Khovanov:2000} will be the presented in a much broader
hashion in a future publication \cite{Kraut:2002}. The reported new
phenomenon is of general relevance for many physical and chemical problems.
It is predicted to occur in a variety of systems, and should experimentally
be observable.
\\
We acknowledge A. Hamm and D. Luchinsky for valuable discussions and
A. Hamm also for the help in programming. This work was supported by the DFG
and INTAS.

\end{document}